\documentclass{article}
\usepackage{amsmath}                             
\usepackage{amsfonts}
\usepackage{epsfig}
\input psfig.sty
\def\fig{.}

\def\n{\noindent}

\begin{document}

\baselineskip .7cm

\author{Navin Khaneja \thanks{To whom correspondence may be addressed. Division of Applied Sciences, Harvard University, Cambridge, MA 02138.
Email:navin@hrl.harvard.edu},\ \ Burkhard Luy,\ \ Steffen J. Glaser \thanks{Institute
of Organic Chemistry and Biochemistry II, Technische Universit\"at M\"unchen,
85747 Garching, Germany.  This work was funded by the Fonds der 
Chemischen Industrie and the Deutsche Forschungsgemeinschaft under 
grant Gl 203/4-1.}}

\title{\bf Boundary of Quantum Evolution under Decoherence}



\maketitle
\begin{center}
{\bf Abstract}
\end{center}
\n 
Relaxation effects impose fundamental limitations on our ability to coherently control quantum
mechanical phenomena.
In this letter, we establish physical
limits on how closely can a quantum mechanical system be steered to a desired target state in the
presence of relaxation. In particular, we explicitly compute the maximum coherence or polarization
that can be transferred between coupled nuclear spins in the presence of very general
decoherence mechanisms that include cross-correlated relaxation. We give analytical expressions
 for the control laws (pulse sequences) which achieve these physical limits and provide supporting
experimental evidence. Exploitation of cross-correlation effects has recently led to the development
of powerful methods in NMR spectroscopy to study very large biomolecules in solution. We demonstrate
with experiments that the optimal pulse sequences provide significant gains over these state of the
art methods, opening new avenues for spectroscopy of much larger proteins. Surprisingly, in spite of
very large relaxation rates, optimal control can transfer coherence  without any
loss when cross-correlated relaxation rates are tuned to auto-correlated relaxation rates.
\vskip 3em

\newpage
\section{Introduction}

The control of quantum ensembles has many applications,
ranging from coherent spectroscopy to quantum information processing. 
In practice, the quantum system of
interest is not isolated but interacts with its environment. This leads to the phenomenon of
relaxation, which results
in  signal loss  and ultimately limits the range of applications.  Relaxation is a major road block
standing in the way of practical quantum computing. Manipulating
quantum systems in a manner that minimizes 
relaxation losses is  a fundamental challenge of
utmost practical importance. 
What is the ultimate limit on how close an ensemble of quantum systems can be steered
from an initial state to a desired target state in the presence of relaxation? 
Until now there existed no theory that answers this question.
This situation is comparable to the time before the fundamental limits of a heat
engine were known:
More than hundred years after the
invention of the steam engine, the physical limits for the maximum amount of work
a steam engine could produce was unclear, in spite of decades of advances in its design. "The theory
of its operation is rudimentary and attempts to improve its performance are still made in an almost
haphazard way"
\cite{Carnot}.  Of course, the maximum efficiency of a heat engine  is
not given by  the cleverness of the engineer who attempts to builds such a machine, but by the
fundamental law of thermodynamics as captured in Carnot's principle.

In this manuscript
we derive fundamental limits on how close can an ensemble of nuclear spins be driven from its initial
state to a desired target state in the presence of relaxation.
In particular, we derive the maximum efficiency of polarization and coherence transfer between
coupled nuclear spins.
A premier example where such coherence transfer operations are important is nuclear
magnetic resonance (NMR) spectroscopy \cite{Ernst}. In structural biology, NMR spectroscopy is an
important technique that allows to determine the structure of biological
macro molecules, such as proteins, in aqueous solution. 
With increasing size of molecules or molecular complexes, the rotational tumbling of
the molecules becomes slower and leads to increased relaxation losses.
When these relaxation rates become comparable to the spin-spin couplings, the efficiency of coherence 
transfer is considerably reduced, leading to poor sensitivity and 
increased measurement times. Recent advances have made it possible to significantly extend the size
limit of biological macro molecules amenable to study by liquid state NMR
[3-6].
These techniques take advantage of the phenomenon of cross-correlated relaxation.
Cross-correlated relaxation represents interference effects between two different ralaxation
mechanisms \cite{Goldman}.
Until now it was not clear if further improvements can
be made and what is the physical limit for the coherence transfer efficiency between coupled spins in
the presence of cross-correlated relaxation.

In this letter, we give analytical expressions for this maximum
achievable coherence transfer efficiency for two heteronuclear coupled spins under
very general decoherence mechanisms that include cross-correlated relaxation. We describe the optimal
pulse sequences that achieve this efficiency and experimental data that
supports these results. We demonstrate that in the limit where the interference effects
become comparable to the uncorrelated relaxation rates, complete coherence transfer is possible
without any loss. In the general case of cross-correlated relaxation, we demonstrate substantial
improvement over previously known sequences in NMR spectroscopy.


\section{Theory}

We consider an isolated heteronuclear spin system
consisting of two coupled spins 1/2, denoted  $I$ (e.g. $^1$H) and $S$ (e.g. $^{15}$N). 
To fix ideas, we first address the problem of 
selective population inversion of two
energy levels (e.g. $\alpha \beta$ and $\beta \beta$) as shown in Fig. 1. 
This is a central step in high-resolution multi-dimensional NMR spectroscopy and
corresponds to the transfer of an initial density operator $I_z$, representing polarization on spin
$I$, to the target state $2I_z S_z$.
\begin{center}
\begin{figure}[h]
\centerline{\psfig{file= \fig/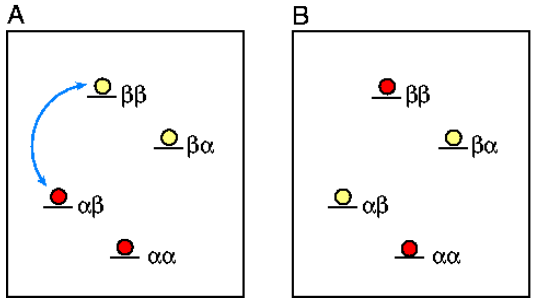 ,width=3.5in}}
\caption{\label{fig:beta_etc} 
 Selective population inversion of the
energy levels $\alpha \beta$ and $\beta \beta$, corresponding to a transfer of polarization
$I_z$ (A) to
$ 2I_z S_z$ (B).} 
\end{figure}
\end{center}
For large molecules in the so-called spin diffusion limit \cite{Ernst}, where longitudinal
relaxation rates are neglegible compared to transverse relaxation rates, both
the initial term ($I_z$) and final term ($2 I_z S_z$) of the density operator are long-lived.
However, the transfer between these two states requires the creation of coherences which in general
are subject to transverse relaxation.
The two principle transverse relaxation mechanisms are dipole-dipole (DD)
relaxation and relaxation due to the chemical shift anisotropy (CSA) of spins $I$ and $S$.
The quantum mechanical equation of motion (Liouville-von Neumann equation) for the
density operator $\rho$ \cite{Ernst} is given by
\begin{eqnarray}\label{eq:rho_dot} 
\nonumber
\dot{\rho} &=& \pi \ J [-i 2  I_z S_z, \rho] 
+ \pi \ k_{DD} [2 I_z S_z, [2 I_z S_z , \rho]
+ \pi \ k^{I}_{CSA} [I_z, [I_z , \rho]]
 + \pi \ k^S_{CSA} [S_z, [S_z , \rho]] \\
 &\ &
+ \ \pi \ k^{I}_{DD/CSA} [2 I_z S_z, [I_z , \rho]] 
 + \pi \ k^{S}_{DD/CSA}  [2 I_z S_z, [S_z , \rho]],
\end{eqnarray}
where $J$ is the heteronuclear coupling constant.
The rates $k_{DD}$, $ k^{I}_{CSA} $, $ k^{S}_{CSA}$ represent auto-relaxation rates due to DD
relaxation, CSA relaxation of spin $I$ and CSA relaxation of spin $S$, respectively. The rates
$k^{I}_{DD/CSA}$ and
$k^{S}_{DD/CSA}$ represent
 cross-correlation rates of spin $I$ and $S$ caused by interference effects between DD and CSA
relaxation. These relaxation
rates
depend on various physical parameters, such as the gyromagnetic ratios of the spins, the
internuclear distance,  the CSA tensors, the strength of the magnetic field and the correlation time
of the molecular tumbling
\cite{Ernst}. Let the initial density operator $\rho(0) = A$ and $\rho(t)$ denote the density operator at time $t$. The maximum efficiency of transfer between $A$ and target operator $C$ is defined as the largest possible value of $trace (C^{\dagger} \rho(t))$ for any time $t$ \cite{Science} (by convention operators A and C are normalized).  

The main result of this letter is as follows.
The maximum efficiency  of transfer between the operators $I_z$ and $2I_z S_z$ depends only on
the scalar coupling constant $J$ and the net auto-correlated and cross-correlated relaxation rates of
spin $I$, given by 
$ 
k_a=k_{DD}+ k^{I}_{CSA} 
$
and 
$
k_c=k^{I}_{DD/CSA}
$, respectively.
This physical limit $\eta$ is given by
\begin{equation}\label{eq:efficiency} 
\eta = 
\sqrt{1+\zeta^2} - \zeta,
\end{equation}  
where
\begin{equation}\label{eq:zeta}
\zeta= \sqrt{{{k_a^2 - k_c^2}\over{J^2 + k_c^2}}}.
\end{equation}
The derivation of the maximum efficiency rests on the basic principles of optimal control theory and
the development of a new class of control systems (see suppl.
material for details). The optimal transfer scheme (CROP: cross-correlated relaxation optimized pulse)
has two constants of motion.  If $l_1(t)$ and $l_2(t)$ denote the two-dimensional vectors
$(\langle I_x \rangle(t),
 \langle I_y\rangle(t))$ and $(\langle 2 I_x S_z \rangle(t),
 \langle 2 I_y S_z \rangle(t))$, respectively, then 
throughout the
transfer process the ratio of the magnitudes of the vectors $l_2$ and $l_1$ is maintained constant at
$\eta$. Furthermore, the angle $\gamma^\ast$ between $l_1$ and $l_2$ is constant throughout.
The two constants of motion of the optimal transfer scheme determine the amplitude and phase
of the rf field at each point in time and explicit expressions for the optimal pulse sequence can be
derived (see suppl. material).

We now consider two important limiting cases of this
problem: 

(I) In the case when  $k_a>0$ and  $k_c=0$ (no
cross-correlated relaxation), the optimal efficiency $\eta$ is equal to $\sqrt{1+ {{k_a^2 
}\over{J^2}}}- {{k_a 
}\over{J}} <1$ (see yellow curves in Figs. 2 and 3) and the optimal angle $\gamma^\ast$
is $\pi/2$ \cite{Rope}.

(II) In the limit where the cross-correlation coefficient
$k_c/k_a$ approaches 1, the optimal transfer efficiency
$\eta$  approaches  1 (see black curves in Figs. 2 and 3) and $\gamma^\ast$ approaches $\pi$. 
Surprisingly, in this case using
optimal control it is possible to transfer coherence without any loss in the presence of relaxation.
In the limit of large relaxation rates $k_a$, this relaxation-optimized transfer mechanism gains up
to 100\% compared to state of the art transfer schemes.

\begin{center}
\begin{figure}[h]
\centerline{\psfig{file= \fig/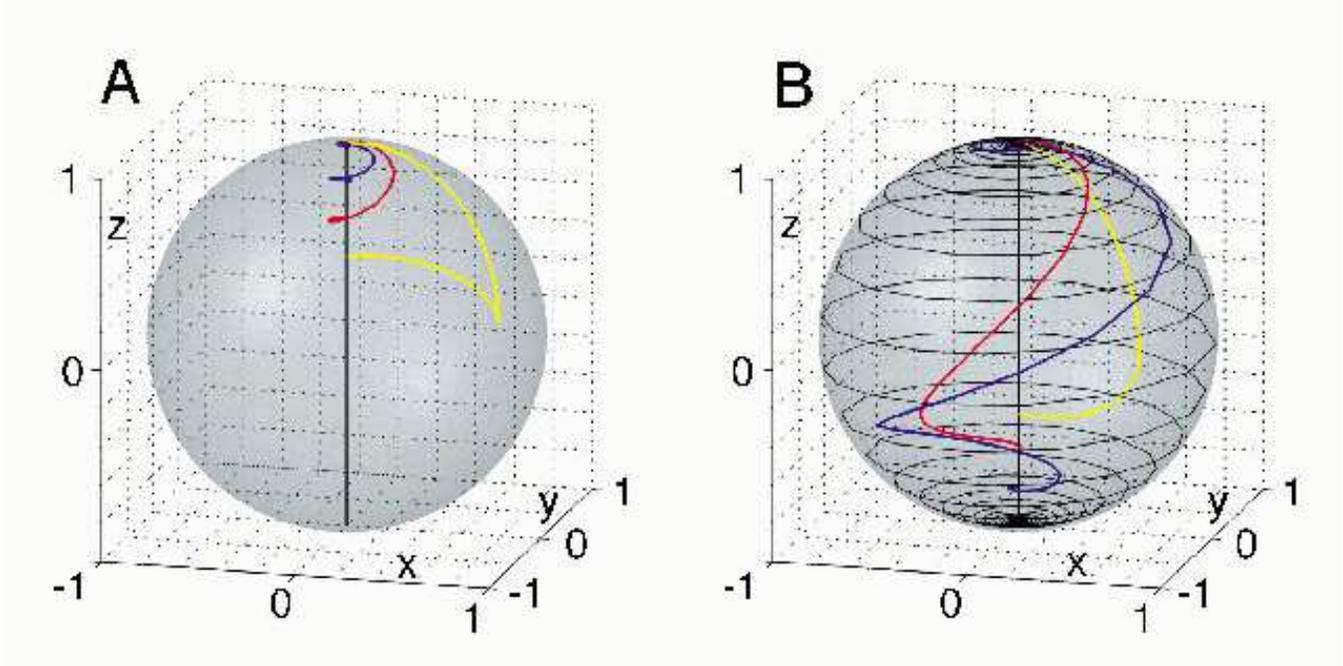 ,width=7in}}
\caption{\label{fig:shere} Optimal trajectories of the two mutliplet components (A)
$\vec{I}S_\alpha$ and (B) $\vec{I}S_\beta$ for $k_a=J$ and $k_c/k_a=0$ (yellow curves),
$k_c/k_a=0.75$ (red curves), $k_c/k_a=0.95$ (blue curves), $k_c/k_a=0.999$ (black
curve).}
\end{figure}
\end{center}

 The optimal transfer scheme is best illustrated by decomposing
the initial operator $I_z$  as a sum of the two operators $I_z
S_\alpha= \frac {I_z}{2} + I_z S_z$ and 
$I_z S_\beta=\frac {I_z}{2} - I_z S_z$. 
The transverse components $I_x S_\alpha$, $I_y S_\alpha$ and $I_x S_\beta$, $I_y S_\beta$ relax with
rates $k_a + k_c$ and
$k_a - k_c$, respectively.
When $k_c/k_a$ approaches 1, the
transverse operators $I_x S_\beta$ and $I_y S_\beta$ do not relax.
The optimal control in this case reduces to selectively inverting $I_z S_\beta$ to $-I_z
S_\beta$ by weak rf irradiation at the frequency ($-J/2$) of the slowly relaxing multiplet component.
Such selective inversions have been performed in the past in the absence of cross-correlated
relaxation \cite{TSI, SPI}. However, since the component $I_z S_\alpha$, which  we do not want to
invert, has a large transverse relaxation rate given by $k_a + k_c$, it is possible to carry out the
selective inversion process
much more rapidly. In Fig. 2, optimal trajectories of the two multiplet components are shown for
several cross-correlation coefficients $k_c/k_a$ and $k_a=J$.

In Fig. 4, the optimal rf amplitude and irradiation
frequency of a CROP sequence is shown as a function of time  for the case $k_c/k_a=0.75$ and
$k_a=J$. Although the ideal sequence has a long duration, most of the transfer occurs in a
relatively short time window, outside of which the rf amplitude is vanishingly small.
The transfer efficiency $\eta$ is shown in
Fig. 3 A for several ratios $k_c/k_a$ as a function of the auto-correlated relaxation rate $k_a/J$.
For the case $k_c/k_a=0.75$, the physical limit of the transfer efficiency
is compared in Fig. 3 B to the transfer efficiency of conventional transfer schemes.

The optimal control methods for the transfer from $I_z$ to $2I_z S_z$ in the presence of
cross-correlated relaxation immediately extend to  other routinely used transfer, such as inphase to
inphase transfer ($I_x \rightarrow S_x $) \cite{ref_INEPT} and single transition to single transition
transfer ($ 2I_x S^\alpha \rightarrow 2 I^\alpha S_x$) \cite{TROSY1}.
Since the operators $I_z$, $S_z$ and $2I_z S_z$ do not decay, the optimal efficiency for the transfer
$I_x$ to
$S_x$ is achieved by first rotating $I_x$ to $I_z$ (which can be done rapidly with neglegible loss).
Then
$I_z$ is transferred optimally to
$2I_z S_z$ with efficiency $\eta$ (Eq. 2), followed by the optimal transfer of
 $2I_z S_z$ to  $S_z$, which is finally rotated rapidly to $S_x$.
The optimal  transfer $2 I_zS_z \rightarrow S_z $
is analogous to the optimal transfer $I_z \rightarrow 2 I_z S_z $. The efficiency $\eta^\prime$ for
this transfer is also given by Eq. (2), where the  rates  $k_a$ and $k_c$ are replaced by the
corresponding rates
$k_a^\prime=k_{DD}+ k^{S}_{CSA}$
and 
$
k^\prime_c=k^{S}_{DD/CSA}$ for spin $S$ and $\zeta$ is replaced by the corresponding  $\zeta^\prime$.
The maximum efficiency for the transfer $I_x \rightarrow S_z $ is the product of the
efficiencies of the individual steps (see Table 1).

\begin{center}
\begin{figure}[h]
\centerline{\psfig{file= \fig/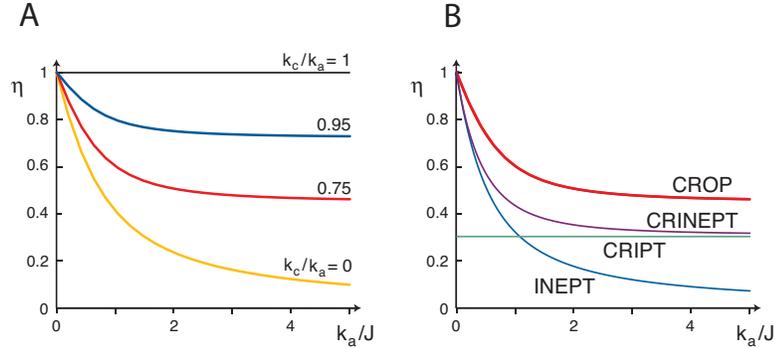 ,width=4 
in}}
\caption{\label{fig:dissipation} (A) Physical limits of the transfer efficiency $\eta$ as a function
of
$k_a/J$ for
$k_c/k_a=0$ (yellow curve),
$k_c/k_a=0.75$ (red curve), $k_c/k_a=0.95$ (blue curve), $k_c/k_a=1$ (black
curve). (B) For the case $k_c/k_a=0.75$, the theoretical bound of the transfer efficiency
(CROP: red curve) is compared to the transfer efficiency of conventional transfer schemes (INEPT:
blue curve, CRIPT: green curve, CRINEPT: purple curve).}
\end{figure}
\end{center}

 \begin{figure}[h]
\centerline{\psfig{file= \fig/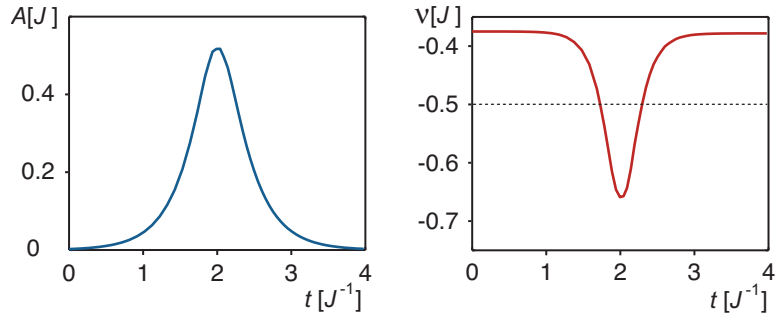 ,width=4
in}}
\caption{\label{fig:rf} 
Truncated cross-correlated relaxation optimized pulse
(CROP) for $k_c/k_a=0.75$  and $k_a=J$:
Radio frequency amplitude $A=\gamma B_0/2 \pi$ (left) and irradiation frequency
$\nu$ (right) as a function of time.}
\end{figure}

In the light of increasing use and superiority of TROSY (Transverse Relaxation-Optimized Spectroscopy)
methods
\cite{TROSY1}, the single transition to single transition transfer $ 2 I_x S^\alpha \rightarrow2
I^\alpha S_x$ is  important in NMR applications to structural biology. It is of both
theoretical and practical interest to establish the physical limits for this transfer. This transfer
can be achieved optimally as a sequence of the following steps. First the term 
$ 2 I_x S^\alpha$ is rapidly rotated to $ 2 I_z S^\alpha= I_z + 2I_z S_z$. In a second step, $2I_z
S_z$ is transferred via CROP to $S_z$, followed by the CROP transfer of $I_z$ to $2 I_z S_z$. This
completes the transfer from $ 2 I_x S^\alpha$ to  $2 I^\alpha S_z$ which is finally rapidly rotated
to  $2 I^\alpha S_x$. The maximum overall transfer efficiency is given by $\sqrt{\eta^2 + {\eta^\prime}^2}$ (c.f. Table 1).

\begin{table*}
\caption{Bounds on Coherence and Polarization Transfer}
\begin{center}
\begin{tabular}{cc} \hline
Transfer&Physical Limits of  Efficiency \\ \hline

$I_z \leftrightarrow 2I_z S_z$ &$\eta=\sqrt{1+\zeta^2} - \zeta$ \\
$2 I_z S_z \leftrightarrow  S_z$ & $eta^\prime=\sqrt{1+{\zeta^\prime}^2} - \zeta^\prime$ \\
$I_z \leftrightarrow S_z$ &   $\eta \eta^\prime$ \\
$I_x S_\alpha \leftrightarrow I_\alpha S_x$ &   $\sqrt{\eta^2 + {\eta^\prime}^2}$ \\
\hline
\end{tabular}
\label{tab:times}
\end{center}
\end{table*}

\section{Experimental results}

The performance of the analytically derived CROP sequences was tested experimentally using the coupled
two-spin system of $^{13}$C-labeled sodium formate with a coupling of $J=193.6$ Hz between the 
$^{13}$C spin (denoted
$I$) and the
$^{1}$H spin (denoted $S$).
In order to control the rotational correlation time, sodium formate was dissolved in a mixture of 96\% 
D$_6$-glycerol and 4\% D$_2$O. The viscosity of this solvent can be conveniently adjusted through a
variation of temperature. 
The experiments were performed at a
temperature of 256.5 K where $k_a/J \approx 1$ (see Fig. 5 A) and 260 K where $k_a/J \approx
0.5$ (see Fig. 5 B).
 At a magnetic
field of 17.6 T, the experimentally determined ratio of
cross and auto correlation rate was $k_c/k_a \approx 0.75$. 
In the preparation phase of the
experiments,  the thermal equilibrium $^1$H
magnetization was dephased by applying a 90$^\circ$ proton pulse followed by a pulsed magnetic field
gradient.
The transfer efficiency of $^{13}$C polarization $I_z$ to $2I_z S_z$ was measured for the novel CROP
sequence, as well as for INEPT \cite{INEPT2}, CRIPT \cite{cript3} and CRINEPT \cite{CRINEPT}
sequences. Finally, a hard 90$_y^\circ$ proton pulse was applied to transform $2I_z S_z$ to  $2I_z
S_x$  and the amplitude of the resulting proton
anti-phase signal was measured. 
The resulting experimental transfer amplitudes are shown in Fig. 5  as a function of the transfer
time. CROP sequences were truncated symmetrically to acquire transfer amplitudes
also for finite mixing times.  Experimentally, the optimal transfer time of the CROP sequence was
found to be 7.5 ms. This is a compromise between losses due to the truncation of the (very
long) CROP sequence and losses due to the non-zero relaxation rates of the terms $I_zS_z$. The
experimentally determined relaxation time of these terms was about 50 ms.
In spite of these non-idealities of the model system, the CROP sequences are substantially more
efficient than the conventional sequences. In Fig. 5 A and B, the experimental gains
compared to CRINEPT are 34\% and 22\%, respectively. We found the although the optimal pulse
sequences were designed for specific rates $k_a$ and $k_c$, they were robust to variations in these
parameters.

\begin{center}
\begin{figure}[h]
\centerline{\psfig{file= \fig/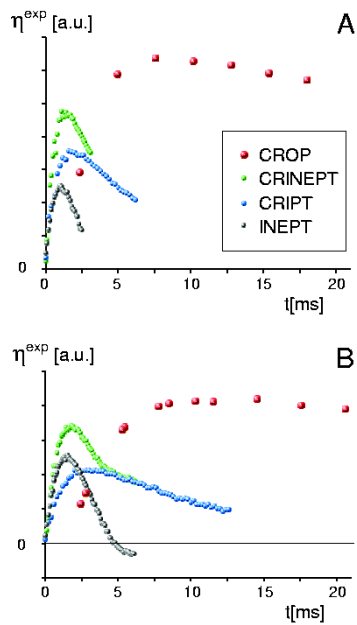 ,width=2.8 
in}}
\caption{\label{fig:dissipation} Experimental transfer amplitude of truncated CROP sequences compared
to
 CRINEPT, CRIPT, and INEPT as a function of total transfer time. The experiments were performed at a
temperature of 256.5 K with $k_a/k_c \approx 1.1$ (A) and 260 K with $k_a/k_c \approx 0.6$(B).}
\end{figure}
\end{center}

\section{Conclusion}

Here, we derived for the first time
upper achievable physical limits on the efficiency of coherence and polarization transfer 
for two coupled
spins in the presence of very general decoherence mechanisms that include
cross-correlated relaxation. 
In this letter, the focus was on the study of 
polarization and coherence transfer between
an  isolated pair  of scalar coupled
heteronuclear spins in the spin diffusion limit.
For this example, new transfer schemes were found which yield substantial gains (of
up to 100\%) in transfer efficiency over conventional methods. 
With these physical limits established, it is expected that significant improvement can be achieved
over state of the art experiments in protein NMR spectroscopy. Work is in progress  to
incorporate practical considerations like
broadbandedness and robustness with respect to variations of relaxation rates and experimental
imperfections. 
The methods presented here can be generalized for finding relaxation optimized pulse sequences in
larger spin systems as commonly encountered e.g. in backbone and side chain assignments in protein NMR
spectroscopy. Furthermore these methods directly extend to other routinely used experiments like
excitation of multiple quantum coherence \cite{Ernst}. The most surprising aspect of the presented
results is that in spite of large relaxation rates, it is possible to exploit the structure of
relaxation and have decoherence-free evolution by steering the system through a decoherence-free
subspace.  It is also expected that the methods presented here will be further developed to minimize
decoherence losses in various proposed implementations of quantum information processing.

\newpage

\begin{center}\section{\bf Supplementary Material: Derivation of Optimal Control of Coherence Transfer under Cross-Correlated Relaxation}\end{center}

The derivation of optimal control assumes that the two heteronuclear spins have  well separated resonance frequencies, allowing for
fast selective manipulation of each  spin on a time-scale determined by the coupling $J$
and the relaxation rates $k_a$ and $k_c$. 
Therefore the
Cartesian spin operator $I_z$
can be
 transformed
to an operator of the form $ I_x \cos \beta_1 + I_z \sin \beta_1$ by the use of strong,
spin-selective radio frequency (rf) pulses without relaxation losses (see Fig. 6). Let $r_1(t)$ represent the magnitude of polarization and in-phase coherence on spin $I$
at any given time $t$, i.e. $r_1^2(t) = \langle I_x\rangle^2 + \langle I_y\rangle^2 + \langle
I_z\rangle^2$ (where e.g.
$\langle I_{x} \rangle = {\rm trace}\{\rho \ I_{x}  \}$ represents the expectation value of
$I_x$).
Let $l_1(t)$ be the magnitude of in-phase coherence on spin $I$, i.e. 
$l_1^2(t) = \langle I_x\rangle^2 + \langle I_y\rangle^2$ 
and $\beta_1=\cos^{-1}{{l_1}\over{r_1}}$
(see Fig. 6).
Using rf fields, we can exactly control  the angle
$\beta_1$.
Hence we can think of $\cos \beta_1$ as a control parameter and denote it by $u_1$ (see Fig. 6).
 Observe that the operator $I_z$ is invariant under the
evolution equation (1), whereas  $I_x$ and $I_y$  evolve under the
$J$ coupling and also relax. 
For example, $I_x$ evolves under the coupling to $2I_y S_z$ with rate $J$ and cross relaxes to 
$-2I_x S_z$
with rate $k_c$. 
In the plane defined by the operators
$2I_x S_z$ and
$2I_yS_z$,  the direction
in which antiphase coherence begins to build up from an initial coherence $I_x$ forms an angle
$$\theta=\tan^{-1}(\frac{J}{-k_c})$$
with the axis $2I_xS_z$.

\begin{center}
\begin{figure}[h]
\centerline{\psfig{file= \fig/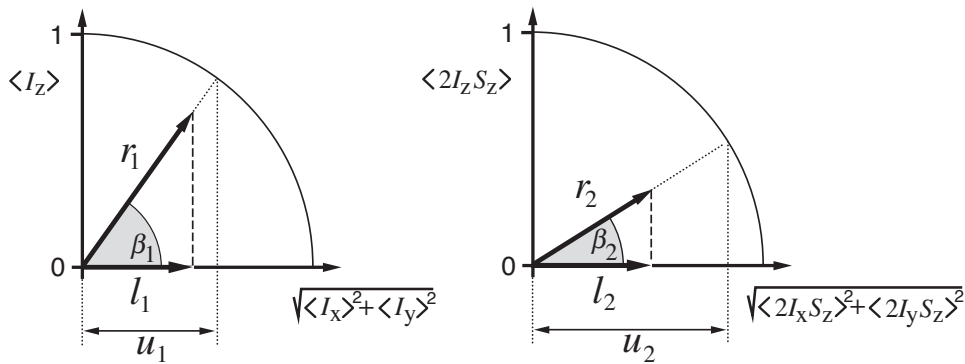 ,width=5 
in}}
\caption{Representation of the system variables $r_1$, $r_2$, their
transverse components
$l_1$,
$l_2$, the angles
$\beta_1$,
$\beta_2$, and of the control parameters
$u_1=\cos 
\beta_1$,
$u_2=\cos \beta_2$ in terms of the expectation values $\langle I_x \rangle$, $\langle I_y
\rangle$, $\langle I_z
\rangle$, $\langle 2 I_x  S_z \rangle$, $\langle 2 I_y  S_z \rangle$, and $\langle 2 I_z  S_z
\rangle$.}
\end{figure}
\end{center}

As the operators $2I_xS_z$ and  $2I_yS_z$ are produced, they also relax. Let $l_2(t)$ measure the magnitude
of the total antiphase coherence at time $t$, i.e.  $l_2^2(t) = \langle
2I_xS_z\rangle^2 + \langle 2I_yS_z\rangle^2 $.
By use of rf pulses it is possible to rotate the antiphase operators
$2I_xS_z$ and $2I_yS_z$ to $2I_z S_z$, which is protected from relaxation. Let $r_2$
represent the total magnitude of the expectation values of these bilinear operators, i.e. $r_2^2(t) =
l_2^2 + \langle 2I_zS_z\rangle^2 $
and $\beta_2=\cos^{-1}{{l_2}\over{r_2}}$
(see Fig. 6).
We can control the angle $\beta_2$ 
 and we define
 $\cos \beta_2$ as a second control parameter $u_2$ (see Fig. 6). 

\begin{center}
\begin{figure}[h]
\centerline{\psfig{file= \fig/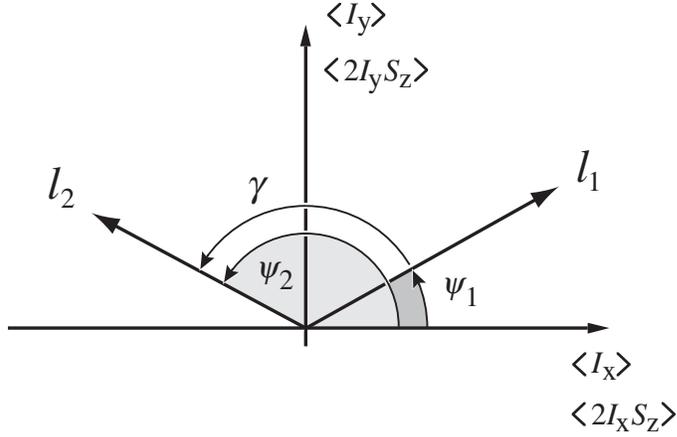 ,width=3.5 
in}}
\caption{The figure shows the transverse planes defined by $I_x$ and $I_y$ superimposed on the plane defined by $2I_xS_z$ and $2I_yS_z$ such that $I_x$ is aligned with $2I_xS_z$. $l_1$ and $l_2$ are the vectors representing 
transverse coherences in these two planes and $\gamma$ is the angle between the vectors.}
\end{figure}
\end{center}

We superimpose the transverse planes defined by $I_x$ and $I_y$ with the plane defined by $2I_xS_z$ and $2I_yS_z$ such that $I_x$ is aligned with $2I_xS_z$ (see Fig. 7). If $\gamma$ represents the angle between $l_1$ and $l_2$, then by definition of $\theta$ we have 
\begin{eqnarray*}
\frac{d}{dt}\ {l_1}(t) &=& - \pi [\ k_a l_1(t) - \sqrt{k_c^2 + J^2} \cos(\theta + \gamma)\ l_2(t)] \\ 
\frac{d}{dt}\ {l_2}(t) &=& - \pi [\ k_a l_2(t) - \sqrt{k_c^2 + J^2} \cos(\theta - \gamma)\ l_1(t)].
\end{eqnarray*} This can be rewritten as 
\begin{equation}\label{eq:for.l}\frac{d}{dt} \left[ \begin{array}{c} l_1(t) \\ l_2(t) 
\end{array} \right] = 
\pi J  \left[ \begin{array}{cc} - \xi  & \chi \ \cos(\theta + \gamma) \\ \chi \ 
\cos(\theta - \gamma) & - \xi 
\end{array}
\right] \left[ \begin{array}{c}
l_1(t) \\ l_2(t) \end{array} \right], \end{equation} where 
$$\xi = k_a/J$$  and 
$$\chi = \sqrt{1+\left({{ k_c}\over{J}}\right)^2}.$$ For a given value of $\beta_1$ and $\beta_2$, we have
$l_1(t) = r_1(t)\cos(\beta_1)$ and $l_2(t) = r_2(t)\cos(\beta_2)$. Since $\frac{d \langle I_z \rangle}{dt} = 0$ and $\frac{d \langle 2 I_z S_z \rangle}{dt} = 0$, we get $\dot r_1(t) = \dot l_1(t)\ \cos(\beta_1)$ and $\dot r_2(t) = \dot l_2(t)\ \cos(\beta_2)$. Substituting for $\dot{l_1}$ and $\dot{l_2}$, we then get 

\begin{equation}\label{eq:main.control}\frac{d}{dt} \left[ \begin{array}{c} r_1(t) \\ r_2(t) 
\end{array} \right] = 
\pi J  \left[ \begin{array}{cc} - \xi u_1^2 & \chi \ u_1 u_2 \cos(\theta + \gamma) \\ \chi \ u_1 u_2
\cos(\theta - \gamma) & - \xi u_2^2
\end{array}
\right] \left[ \begin{array}{c}
r_1(t) \\ r_2(t) \end{array} \right]. \end{equation} 

Given the dynamical system in equation
(\ref{eq:main.control}), we want to find the optimal values of  $u_1(t)$, $u_2(t)$, and $\gamma(t)$, so that
 starting from $r_1(0) = 1$ we achieve the largest value for $r_2$. 

As the operator $I_z$ is transferred to $2I_zS_z$, the ratio $\frac{r_2}{r_1}$ increases from $0$ to $\infty$. The optimal choice of $u_1$ and $u_2$ must ensure that the ratio of gain $d r_2$ in $r_2$ to loss $d r_1$ in $r_1$ for incremental time steps $dt$ is maximized at each step. This ratio is 
$$ \frac{|\dot {r_2}|}{|\dot {r_1}|} =  \frac{ - \xi u_2^2 r_2 + \chi u_1u_2 r_1 \cos(\theta - \gamma) } {\xi u_1^2 r_1 - \chi u_1 u_2 \cos(\theta + \gamma) r_2 }.$$ Let $\frac{u_2 r_2}{u_1 r_1} = g$, then the above expression can be re-written as 
$$ \frac{|\dot{r_2}|}{|\dot{r_1}|} =  \frac{r_1}{r_2} \ \frac{- \xi g^2 + \chi\ g\ \cos(\theta - \gamma) }{\xi - \chi\ g\ cos(\theta + \gamma)}. $$ This expression needs to be maximized with respect to choice of $g$ and $\gamma$. Let these optimal values be $\eta$ and $\gamma^{\ast}$ respectively. Then $$ \frac{d \frac{|\dot{r_2}|}{|\dot{r_1}|}}{dg}|_{g = \eta, \gamma = \gamma^{\ast}} = 0 $$ yields
\begin{equation} \label{eq:variation1} \frac{1}{\eta} \cos(\theta - \gamma^{\ast}) + \eta \cos(\theta + \gamma^{\ast} ) = \frac{2 \xi}{\chi}, \end{equation} which yields 
$ \frac{|\dot{r_2}|}{|\dot{r_1}|} =  \frac{r_1}{r_2} \eta^2$. Now the value of $\gamma^{\ast}$ in equation (\ref{eq:variation1}) is such that it maximizes $\eta$. Differentiating both sides of equation (\ref{eq:variation1}) with respect to $\gamma^{\ast}$ and substituting $\frac{d \eta}{d \gamma^{\ast}}= 0$, we obtain that 
\begin{equation}\label{eq:variation2} \frac{1}{\eta} \sin(\theta - \gamma^{\ast} ) - \eta \sin(\theta + \gamma^{\ast} ) = 0. \end{equation} The optimal $\eta$ and $\gamma^{\ast}$ then satisfy equations (\ref{eq:variation1}, \ref{eq:variation2}). The two equations can then be solved to give \begin{equation}\label{eq:efficiencysup} \eta = \sqrt{\zeta^2 + 1} - \zeta, \end{equation} where $\zeta = \sqrt{{k_a^2 - k_c^2}\over{J^2 + k_c^2}}$ and optimal $\gamma^{\ast} =\tan^{-1} {{1-\eta^2}\over{(1+\eta^2)\cot \theta}}$. By substituting the optimal control law $\frac{u_2(t)}{u_1(t)} = \frac{\eta r_1(t)}{r_2(t)}$, and integrating equation (\ref{eq:main.control}), we see that $r_2$ increases from $0$ to $\sqrt{\zeta^2 + 1} - \zeta $, which is then the maximum achievable transfer.

\noindent We see that throughout the optimal transfer, the angle $\gamma$ is maintained constant at $\gamma^\ast$. The optimal control ensures that the ratio of the transverse components $l_2(t)$ and
$l_1(t)$ is always maintained constant at $\eta$. These two constraints can now be used to get explicit expressions for the magnitude and phase of the optimal rf-field.

Let $\phi$ denote the phase of the rf-field relative to $l_1$ (\ $\gamma - \phi$ relative to $l_2$ ) and let $A$ be its amplitude. Let $dl_1^{\perp}$ denote the change in transverse component perpendicular to the vector $l_1$ by application of the $rf$ field in small time $dt$. Then observe $dl_1^{\perp} = 2 \pi A \langle I_z \rangle \cos(\phi) dt$. Similarly  $dl_2^{\perp} = 2 \pi A \langle 2 I_z S_z \rangle \cos(\gamma -\phi) dt$. If $\frac{l_2}{l_1}$ is maintained at $\eta$ then the angle $\gamma$ does not change due to the evolution equation (\ref{eq:for.l}). Therefore we only need to consider the change in $\gamma$, due to the rf field. If $\gamma$ is maintained constant, then $\frac{dl_1^{\perp}}{l_1} = \frac{dl_2^{\perp}}{l_2}$ . This gives $ \tan(\beta_1) \cos(\phi) = \tan(\beta_2)\cos(\gamma^{\ast} -\phi)$ because $\frac{\langle I_z \rangle}{l_1} = \tan \beta_1$ and $\frac{\langle 2 I_z S_z \rangle}{l_2} = \tan \beta_2$.  This then implies that
$$ \phi= \tan^{-1} \left( {{\tan \beta_1}\over{\tan \beta_2 \sin \gamma^\ast}} - \cot \gamma^\ast\right). $$ 

The amplitude can be determined from the condition that $\frac{l_2(t)}{l_1(t)}$ is maintained constant. This implies that $\frac{dl_1}{l_1} = \frac{dl_2}{l_2}$. Substituting $dl_1 = ( -\xi \pi J l_1 - \chi \pi J \cos(\theta + \gamma) l_2 + 2 \pi A \langle I_z \rangle \sin(\phi) \ ) dt$ and $dl_2 = ( -\xi \pi J l_2 - \chi \pi J \cos(\theta - \gamma) l_2 - 2 \pi A \langle 2 I_z S_z \rangle \sin(\gamma - \phi)\ ) dt$, we get 
$$ A= \frac{1}{2 \pi} \ {{\left( \cos(\theta-\gamma^\ast) - \eta^2 \cos(\theta+ \gamma^\ast)\right) \chi J }\over {\left(\tan \beta_1 \sin \phi + \tan \beta_2 \sin(\gamma^\ast-\phi) \right)\eta}}.$$

The expressions of $A$ and $\phi$ are given in terms of the state of the system (angle $\beta_1$ and $\beta_2$). We can insert these in the equations for how $\beta_1$ and $\beta_2$ evolve as a function of $A$ and $\phi$ to get explicit expressions of $\beta_1$, $\beta_2$, $\phi$ and $A$ as a function of time. This gives us the amplitude and phase of the optimal rf pulse as a function of time.

\end{document}